\documentclass[%
 ams,
prl,%
 amsmath,amssymb,
 reprint,%
]{revtex4-1}

\usepackage{graphicx}
\usepackage{dcolumn}
\usepackage{bm}

\begin{document}
\begin{abstract}

The combination of electronic correlation and spin-orbit coupling is thought to precipitate a variety of highly unusual electronic phases in solids, including topological and quantum spin liquid states. We report a Raman scattering study that provides evidence for unconventional excitations in $\alpha$-RuCl$_3$, a spin-orbit coupled Mott insulator on the honeycomb lattice. In particular, our measurements reveal unusual magnetic scattering, typified by a broad continuum. The temperature dependence of this continuum is evident over a large scale compared to the magnetic ordering temperature, suggestive of frustrated magnetic interactions. This is confirmed through an analysis of the phonon linewidths, which show a related anomaly due to spin-phonon coupling. These observations are in line with theoretical expectations for the Heisenberg-Kitaev model and suggest that $\alpha$-RuCl$_3$ may be close to a quantum spin liquid ground state. 

\end{abstract}

\title{Scattering Continuum and Possible Fractionalized Excitations in $\alpha$-RuCl$_3$}
\author{Luke J. Sandilands}
\author{Yao Tian}
\author{Kemp W. Plumb}
\author{Young-June Kim}
\affiliation{Department of Physics, University of Toronto, 60 St. George St., Toronto, Ontario M5S 1A7, Canada\\}%

\author{Kenneth S. Burch}
\affiliation{Department of Physics, Boston College, 140 Commonwealth Ave., Chestnut Hill, Massachusetts 02467, USA\\}%

\maketitle 

The quantum spin liquid, where quantum fluctuations obstruct long-range order at low temperatures, is one of the most elusive and intriguing states of matter\cite{Balents:2010fk,frustated_book}. These states have been suggested to play a crucial role in high-temperature superconductivity\cite{RevModPhys.78.17} and could lead to topologically protected excitations useful for quantum computation\cite{RevModPhys.80.1083}. The elementary excitations of these systems are fractionalized, reflecting the peculiar nature of the quantum spin liquid (QSL) ground state, and can be identified in scattering experiments where they manifest themselves as broad continua, in contrast to the sharp magnon modes characteristic of ordered magnets\cite{Han:2012fk,PhysRevB.82.144412,PhysRevB.68.134424,PhysRevLett.113.187201}. While QSLs have been extensively studied theoretically, the experimental picture is limited due to the small number of candidate material systems\cite{Balents:2010fk}.

In this context, it was recently suggested\cite{PhysRevLett.102.017205,PhysRevLett.105.027204} that strongly spin-orbit coupled Mott insulators on the honeycomb lattice might manifest a two-dimensional QSL ground state. In such a material, strong spin-orbit coupling can lead to the formation of $j_{eff} = 1/2$ pseudospins\cite{PhysRevLett.101.076402,PhysRevLett.102.017205,PhysRevLett.105.027204}, local moments of mixed spin and orbital character, whose bond-dependent exchange interactions map onto the Kitaev Hamiltonian: $H_K = -J_{K}\sum_{\gamma-links}S_i^\gamma S_j^\gamma$. Here $S_i^\gamma$ refers to the $\gamma$ component of the spin at the $i^{th}$ lattice site and the sums are performed over the three different types of links of the honeycomb lattice $\gamma = {x,y,z}$. For a given site, the exchange interactions along all three links cannot be simultaneously satisfied. These bond-dependent interactions are therefore intrinsically frustrated, in a manner distinct from geometric frustration, and lead to an exotic, spin-disordered ground state and fractionalized Majorana fermion excitations\cite{Kitaev20062}. Experimental efforts in this direction have so far focused on the honeycomb lattice iridates\cite{PhysRevB.82.064412,PhysRevB.87.220407,PhysRevLett.110.076402}, although these compounds are structurally complex \cite{PhysRevLett.108.127204,PhysRevB.85.180403} and the applicability of Kitaev physics has also been questioned\cite{PhysRevLett.109.197201,PhysRevB.88.035115,PhysRevB.85.180403}. 

\begin{figure}
  \includegraphics[width=\columnwidth]{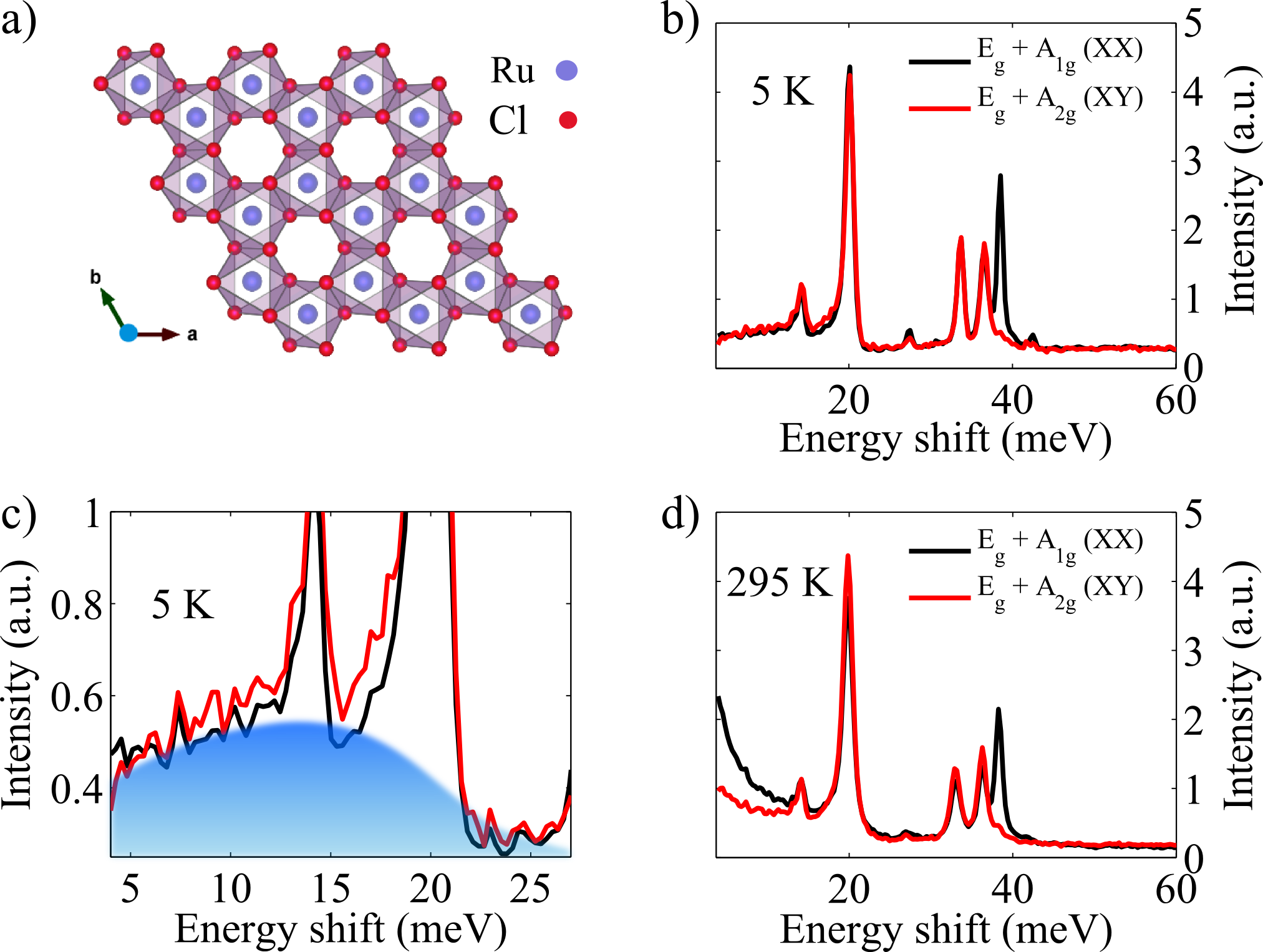}%
\caption{\label{fig:1}Structure and polarized Raman response of $\alpha$-RuCl$_{3}$. (a) Lattice structure of a single plane of $\alpha$-RuCl$_3$. The Ru atoms are shown in blue while the Cl are indicated in red. (b) 5~K Raman intensity. (c) Low-energy detail at 5~K. The shaded blue region is a guide to the eye and indicates the continuum contribution. (d) 295~K Raman intensity. The Raman spectra evince a series of narrow phonon modes in both channels, a broad continuum extending up to 20 meV in $E_g$, and a quasielastic scattering component in $A_{1g}$. The continuum emerges at low temperatures, while the quasielastic scattering is only visible at high temperatures.}%
 \end{figure}

To experimentally explore spin liquid behavior driven by spin-orbit coupling, we have investigated the elementary excitations of $\alpha$-RuCl$_3$ using polarized Raman scattering. $\alpha$-RuCl$_3$ is a recently identified spin-orbit assisted Mott insulator\cite{PhysRevB.90.041112} which crystallizes in a layered structure with planes of edge-sharing RuCl$_6$ octahedra arranged in a honeycomb lattice\cite{Stroganov1957} [Fig. \ref{fig:1} (a)] and is therefore an excellent candidate for realizing the Heisenberg-Kitaev model in the solid state\cite{PhysRevLett.102.017205,PhysRevLett.105.027204}. The magnetic properties of $\alpha$-RuCl$_3$ have been subject to a number of previous studies\cite{Stroganov1957,doi:10.1021/ic00048a025,J19670001038}. Magnetic susceptibility and specific heat measurements suggest a pair of magnetic phase transitions near 8 K and 14 K\cite{2014arXiv1411.6515M,2014arXiv1411.4610S}. Intriguingly, a recent single crystal neutron diffraction experiment suggests a zigzag or stripy-type magnetic ordering with a small ordered moment\cite{2014arXiv1411.4610S}, as expected for a Heisenberg-Kitaev magnet close to a QSL ground state\cite{2014arXiv1411.6623V,PhysRevLett.112.077204}. However, no spectroscopic information exists regarding the spin excitations of $\alpha$-RuCl$_3$ and the possibility of spin fractionalization driven by spin-orbit coupling, a defining feature of the Kitaev QSL state, therefore remains an open question. Raman spectroscopy is an excellent tool for studying this possibility, as it can simultaneously probe spin, lattice, and electronic excitations\cite{Lemmens20031,Blumberg21111997}.  
 
Our measurements reveal anomalous magnetic light scattering in $\alpha$-RuCl$_3$, typified by a broad continuum below 100~K that persists in the magnetically ordered state(s) below 14~K. This behavior is not easily explained by conventional two-magnon scattering or structural disorder, but is consistent with theoretical predictions for the Kitaev spin liquid\cite{PhysRevLett.113.187201}. These results suggest that $\alpha$-RuCl$_3$ hosts unusual magnetic excitations and may be proximate to a QSL ground state. 

The $\alpha$-RuCl$_3$ crystals used in this study were grown by vacuum sublimation of prereacted RuCl$_3$ powder\cite{PhysRevB.90.041112}. The Raman spectra were measured in the quasi-backscattering geometry in both collinear (XX) and crossed (XY) polarizations, with light polarized in the basal (cleavage) plane. Light from a 532 nm laser was focused down to a 10~$\mu$m spot and the power at the sample is estimated to be 500 $\mu$W. Two holographic notch filters (Ondax) were used to reject light from the fundamental, leading to a lower cutoff of 2 meV in XY and 3.5 meV in XX. The resolution of our spectrometer is estimated to be 0.6 meV. No anti-Stokes scattering was observed at 5~K, suggesting negligible laser-induced heating. This was confirmed by halving the power and doubling the integration time, which produced no noticeable change in the spectra recorded at 5 K.

The polarized Raman intensity of $\alpha$-RuCl$_3$ is shown in Fig. \ref{fig:1} for 5 and 295~K. The 5~K spectrum in Fig. \ref{fig:1} (b) consists of a series of sharp phonon modes superimposed on a weak continuum that extends up to roughly 20 meV. The continuum is shown in detail in Fig. \ref{fig:1} (c). Strong and sharp phonon modes are resolved at 14, 20, 34, 37, and 38 meV, with weaker modes detected at 27 and 42 meV. Due to the broad line shape and low energy, we attribute the continuum to magnetic scattering. In comparison, an electronic mechanism for the continuum is unlikely as $\alpha$-RuCl$_3$ is electrically insulating with a transport gap of 0.3 eV \cite{Rojas1983349}. As we describe later, the magnetic assignment is further justified by the temperature dependence of the continuum scattering, which is inconsistent with a phonon mechanism.  

The phonon spectra can be understood in terms of an isolated layer of $D_{3d}$ symmetry, as anticipated from the layered crystal structure(see the Supplemental Material\cite{supplemental}. The observed modes are sharp, suggestive of a well-ordered crystal lattice, and show no evidence for structural transitions down to 5~K. We note that the two lowest energy phonons of E$_g$ symmetry near 14 and 20 meV show Fano line shapes at low temperatures\cite{PhysRev.124.1866}. This effect arises generically when a narrow resonance couples to a continuum. Furthermore, the Fano line shape of the 20~meV phonon demonstrates that the continuum extends up to $\sim$20$-$25 meV.

 \begin{figure*}
 \includegraphics[scale=0.85]{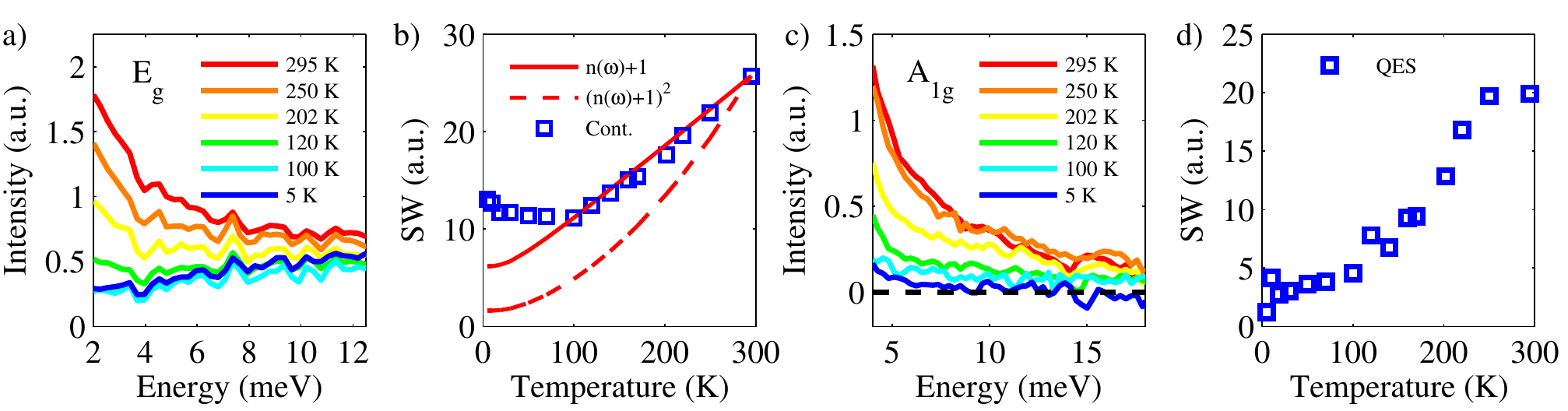}%
 \caption{\label{fig:2}Magnetic scattering in RuCl$_{3}$. (a) Raman intensity in $E_g$ and $A_{2g}$ (XY). (b) $E_g$ and $A_{2g}$ (XY) low-energy spectral weight (SW) vs. temperature. (c) Raman intensity in $A_{1g}$ (XX-XY).  (d)   $A_{1g}$ (XX-XY) low-energy spectral weight (SW) vs. temperature. As shown in (c), the $A_{1g}$ quasielastic scattering (QES) decreases in intensity as temperature is reduced down to 100~K. In contrast, the intensity of the $E_g$ continuum decreases as temperature is reduced to 100~K before increasing. This is more clearly resolved in the spectral weights shown in b) and d). The behavior of the $E_g$ continuum is at odds with the temperature dependence of the thermal factors expected for one [$n(\omega)+1$] and two [$(n(\omega)+1)^2$] particle scattering and signals a change in the magnetic excitations near 100~K.}%
 \end{figure*} 
 
Having identified the excitations present, we focus on the temperature dependence of the magnetic scattering. For conventional (e.g. phonon) scattering, the Raman intensity should show a monotonic decrease given by the thermal Bose factor, although this is not the case in $\alpha$-RuCl$_3$.  In Fig. \ref{fig:2} (a), we plot the $E_g$ Raman intensity at several temperatures. The continuum loses intensity before reaching a minimum near 100~K. Below 100~K, the continuum gains intensity without a change in energy scale or shape, in contrast to conventional scattering. This temperature dependence can be appreciated by considering the spectral weight SW = $\int_{\omega_l}^{\omega_h}$ I$(\omega)d\omega$, where the cutoffs $\omega_l$ and $\omega_h$ are 2.5 and 12.5 meV, shown in Fig. \ref{fig:2} (b).  SW decreases before increasing below 100~K, at odds with the trend expected for one- or two-particle scattering due to the thermal Bose factors \{$[n(\omega)+1]$ or $[n(\omega)+1]^2$\}(See the Supplemental Material\cite{supplemental}). The change in the magnetic scattering near 100~K is consistent with the suggested onset of in-plane spin correlations, first noted by Kobayashi $et$ $al.$\cite{doi:10.1021/ic00048a025} in relation to magnetization measurements. Importantly, the persistence of magnetic scattering far above $T_N$ is a signature of frustrated, low dimensional magnetism.

A comparison of the 100~K and 5~K curves in Fig. \ref{fig:2} (a) reveals that the continuum shape does not vary strongly, even below the ordering temperatures of 8 and 14 K. Typically, the broad continuua observed in the paramagnetic state of an antiferromagnet will evolve into well-defined peaks due to one or two magnon excitations\cite{Lake:2005aa}. In contrast, the scattering in $\alpha$-RuCl$_3$ displays no obvious changes upon ordering. We suspect that lower energies or temperatures than are accessible in our experiment may be needed to detect the magnons associated with the ordered magnetic state.

The observed magnetic continuum is unusual and its origin is not immediately clear. X-ray absorption and $ab$ $initio$ studies have firmly established the relevance of $j_{eff} = 1/2$ physics\cite{PhysRevB.90.041112,2014arXiv1411.6623V}, while static probes of the magnetism have hinted at the existence of the Kitaev interaction in $\alpha$-RuCl$_3$\cite{2014arXiv1411.6515M,2014arXiv1411.4610S}. Scattering from fractionalized spin excitations is therefore an appealing interpretation for the magnetic continuum. Indeed, a recent theoretical work identified a broad $E_g$ continuum, extending up to 3$J_K$ (where $J_K$ is the Kitaev interaction as defined earlier), as the Raman signature of the Kitaev spin liquid \cite{PhysRevLett.113.187201}. Equating 3$J_K$ with the experimental upper cutoff of continuum scattering of 20$-$25 meV yields $J_K \sim$ 8 meV. On a more general, empirical level, the broad line shape and temperature dependence of the continuum in $\alpha$-RuCl$_3$ are reminiscent of the behavior found in the putative spin liquid materials herbertsmithite\cite{PhysRevB.82.144412,PhysRevLett.111.127401} and Cs$_2$CuCl$_4$\cite{PhysRevB.68.134424}\footnote{The proximity to a QSL ground state in Cs$_2$CuCl$_4$ has been questioned, see reference \cite{Balents:2010fk}}. In these compounds, the continuum scattering gains intensity as spin liquid correlations develop. Cs$_2$CuCl$_4$ also orders at low temperatures. Similar to $\alpha$-RuCl$_3$, the continuum scattering in Cs$_2$CuCl$_4$ displays a minimal change upon entering the ordered state, with well-defined magnons only evident at very low energies and involving a small portion of the total spectral weight\cite{PhysRevB.68.134424}. Finally, a recent Raman study of (Na$_{1-x}$Li$_x$)$_2$IrO$_3$ also reported a broad continuum that was interpreted in terms of Kitaev QSL physics\cite{2014arXiv1408.2239N}.

A second parallel between $\alpha$-RuCl$_3$ and herbertsmithite is the presence of quasielastic scattering (QES)\cite{PhysRevB.82.144412}. As can be seen in Fig. \ref{fig:1} (d), a QES component emerges in $\alpha$-RuCl$_3$ at high temperatures in the $A_{1g}$ channel. Such scattering is observed in a variety of low-dimensional magnetic systems \cite{Lemmens20031} and is usually assigned to fluctuations in the magnetic energy density. In Fig. \ref{fig:2} (c), we plot the $A_{1g}$ intensity $I_{A_{1g}} \sim I_{xx} - I_{xy}$ for select temperatures. This QES intensity decreases down to 100~K, signifying a gradual evolution of the low-energy spin dynamics, at which point it becomes difficult to resolve in our data. This trend is borne out in the $A_{1g}$ spectral weight, integrated from 4 to 17 meV, shown in Fig. \ref{fig:2} (d)\footnote{The difference in the energy ranges used to evaluate the continuum and QES spectral weights is due to lower experimental cutoff in the $E_{g} + A_{2g}$ data.}. This could be due to a reduction in the intensity or scattering rate of the quasielastic fluctuations, but we cannot discriminate between these two possibilities with our experimental cutoff. Similar to the continuum scattering, the QES displays a strong temperature dependence far above the ordering temperature, as expected for a low-dimensional, frustrated magnet\cite{Lemmens20031}.

\begin{figure}
\includegraphics[width=\columnwidth]{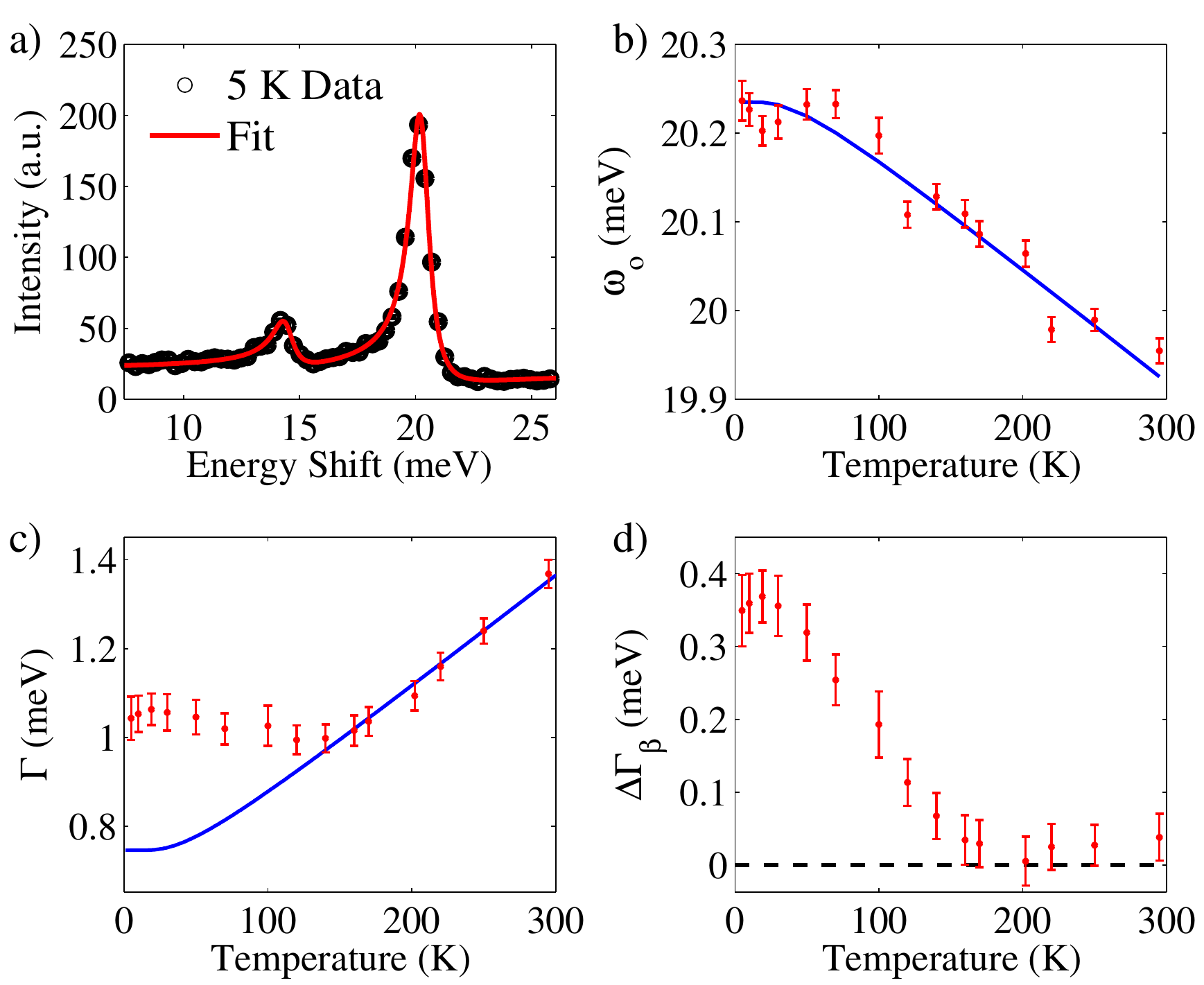}%
 \caption{\label{fig:4} Spin-phonon coupling in $\alpha$-RuCl$_3$. a) Data and Fano fit to the low-energy scattering at 5~K. b) Energy $\omega_o$ of the 20 meV phonon. c) Linewidth $\Gamma$ of the 20 meV phonon. d) Magnetic contribution $\Delta\Gamma$ to the phonon linewidth. The low-energy data are well-described by the Fano form and indicate a coupling between the magnetic continuum and the lattice. This coupling is also apparent in the temperature dependence of $\Gamma$, which shows an anomaly near 140~K.}%
 \end{figure} 

In systems with moderate spin-phonon coupling, the lattice dynamics provide an additional probe of the magnetic degrees of freedom. Indeed, the Fano line shape of the $E_g$ phonons is a signature of spin-phonon coupling in $\alpha$-RuCl$_3$. The spectral region depicted in Fig. \ref{fig:4} (a) is well fitted at 5~K by a function consisting of the sum of two Fano line shapes and a constant background term (See Supplemental Material\cite{supplemental}). In Figs. \ref{fig:4} (b) and (c) we show the linewidth ($\Gamma$) and mode frequency ($\omega_o$) for the 20 meV phonon. As shown in Fig. \ref{fig:4} (c), the linewidth $\Gamma$(T) displays a nonmonotonic temperature dependence. It decreases almost linearly with reduced temperature from 300~K down to 140~K. The width then begins to grow before plateauing at low temperature. 

Phonon self-energies are typically determined by lattice anharmonicity, leading to a monotonic temperature dependence\cite{PhysRevB.29.2051}. However, $\Gamma$(T) reveals a nonmonotonic behavior. To demonstrate this, we have included in Figs. \ref{fig:4} (b) and (c) plots of the behavior expected from anharmonicity (See Supplemental Material\cite{supplemental}). $\Gamma$(T) and the fit diverge near 140~K, implying an additional relaxation mechanism. This is seen in Fig. \ref{fig:4} (d), where we plot $\Delta\Gamma = \Gamma-\Gamma_{anh}$, which reveals an onset near 140~K, similar to the 100~K scale associated with anomalies in the continuum scattering and magnetic susceptibility\cite{doi:10.1021/ic00048a025}. We conclude that there are two contributions to the low-lying $E_g$ phonon self-energies: an anharmonic term and a component due to spin-phonon scattering which onsets near 140~K. This results in an increase in $\Gamma$ and to the Fano line shape of the $E_g$ phonons. The phonon properties shown in Fig. \ref{fig:4} do not show obvious changes in the magnetically ordered phases below 14 K, suggesting that  any effects due to spin-phonon coupling upon ordering are subtle. 


Overall, our experimental results appear consistent with the notion that $\alpha$-RuCl$_3$ hosts unusual magnetic excitations, possibly driven by proximity to a Kitaev QSL state. At this point, however, we consider other possible interpretations of the continuum, namely heavily phonon-damped two-magnon excitations and stacking disorder. Spin-phonon coupling is evident in the data and suggests an alternative explanation for the magnetic continuum: conventional magnons that are heavily damped by magnon-phonon scattering. Indeed, for $D_{3d}$ symmetry, two-magnon (2M) scattering is also expected in the $E_g$ channel\cite{PhysRevB.49.4352}. Additionally, magnon-phonon coupling is known to cause unusually broad 2M features in some cases, for instance in NiPS$_3$\cite{PhysRevB.49.4352} and the insulating cuprates\cite{PhysRevB.42.4842}.  However, we regard the scenario of heavily damped 2M scattering as unlikely in $\alpha$-RuCl$_3$ for several reasons. First, the line shape of the magnetic scattering does not change appreciably when going from 100 to 5~K, in contrast to previous reports of phonon-damped 2M scattering\cite{PhysRevB.49.4352,PhysRevB.42.4842}. For dominant magnon-phonon scattering, the decay rate of a magnon of energy $\omega$ should be roughly proportional to the thermal occupation factor $n(\omega)$+1 of a phonon of the same energy\cite{PhysRevB.42.4842}. Taking the magnon energy to be 10 meV (the center of the continuum), this would imply a $20\%$ change in lifetime at low temperature. For a 5 meV magnon, this change would be $40\%$. Experimentally, however, no significant change is observed in the continuum scattering over this temperature range. Similarly, the characteristic energy of 2M scattering typically blueshifts considerably as temperature is lowered and evolves into a well-defined peak below $T_N$\cite{PhysRevB.49.4352}, in contrast to our observations in $\alpha$-RuCl$_3$ [Fig. \ref{fig:2} (b)]. We therefore conclude that spin-phonon coupling does not appreciably damp the magnetic excitations in $\alpha$-RuCl$_3$. This is a result of the relatively low energy scale of the magnetic excitations compared to the phonons. In compounds where the magnons are heavily phonon damped, the characteristic energies of the magnetic excitations are high compared to the phonon energies, and so a large number of phonon branches are able to participate in the decay process\cite{PhysRevB.49.4352}. In contrast, the magnetic excitations in $\alpha$-RuCl$_3$ are far less energetic than the majority of phonon branches present, meaning the phase space for spin-phonon scattering is significantly reduced.

A further issue is the possible influence of stacking faults. $\alpha$-RuCl$_3$ is a layered material and so such defects are likely present. Indeed, a short out of plane ($c$ axis) magnetic correlation length was reported in a recent neutron diffraction study and attributed to stacking faults\cite{2014arXiv1411.4610S}. This stacking disorder in fact suggests another interpretation of the low-energy continuum reported here. In Raman scattering, stacking faults can lead to a relaxation of the $k=0$ selection rule. In heavily disordered SiC, for example, this leads to both a broadening of existing Raman features and the appearance of new ones, including broad low-energy scattering corresponding to defect-activated acoustic phonon modes\cite{doi:10.1080/01418639408240266}. It is not clear, however, how static structural disorder could produce the unusual temperature dependences and QES present in our data. In the same vein, one might expect a broadening of the magnetic excitations due to stacking faults on the order of the interplane exchange interactions. Since the electronic structure of $\alpha$-RuCl$_3$ is highly two-dimensional\cite{PhysRevB.90.041112}, we expect that the interplane exchange interactions, and thus any broadening due to stacking faults, to be small compared to the overall energy scale of the magnetic continuum, which is set by intraplane interactions.

To conclude, we have studied the elementary excitations of $\alpha$-RuCl$_3$ using Raman scattering. Our measurements reveal a continuum of scattering that is not readily explained by conventional 2M scattering or structural disorder. Rather, the broad continuum reported here appears consistent with theoretical expectations for the Kitaev spin liquid. However, we stress that a categorical assignment is not currently possible and will likely require inelastic neutron scattering. We observe changes in the magnetic dynamics (as evinced by the magnetic continuum and QES, as well as the 20 meV phonon self-energy) near 100$-$140~K, which we interpret as the development of in-plane spin correlations. Overall, our study suggests that $\alpha$-RuCl$_3$ may be close to a Kitaev spin liquid state, consistent with recent neutron diffraction and theoretical studies\cite{2014arXiv1411.4610S,2014arXiv1411.6623V}. 

\begin{acknowledgments}
\end{acknowledgments}

%


%




\clearpage
\section{\textbf{Supplemental Material}}

%

\maketitle

\setcounter{figure}{0}
\makeatletter 
\renewcommand{\thefigure}{S\@arabic\c@figure}
\makeatother

\subsection{I. Phonon Mode Assignment}
Structurally, $\alpha$-RuCl$_3$ consists of a stack of weakly van der Waals bound layers. Accordingly, the number and symmetry of the phonons modes can be understood in terms of an isolated RuCl$_3$ layer of symmetry $D_{3d}$, rather than the full space group. In this case, group theory predicts a total of six (4$E_g$ and 2$A_{1g}$) Raman-active modes\cite{Bermudez1976693}. The 4$E_g$ modes are expected to be visible in both polarization channels while the 2$A_{1g}$ modes vanish in the crossed geometry. Indeed, the modes at 38 and 42 meV are suppressed in the crossed (XY) geometry and so can be assigned A$_{1g}$ symmetry. The remaining five modes as well as the continuum are present in both polarization channels and therefore have E$_g$ symmetry, meaning RuCl$_3$ exhibits one more E$_g$ phonon mode than expected from group theory for a single D$_{3d}$ layer. We therefore suspect that the mode at 27 meV is either defect activated or due to interlayer interactions, as it is the weakest peak in the E$_g$ channel. Given that the spectra show only weak deviations from the ideal $D_{3d}$ symmetry of a single layer, we conclude that interlayer lattice interactions are  indeed weak in this compound and so do not significantly affect the phonon dynamics. Although all observed phonon modes broaden in going from 5 to 295~K, the number of modes does not change, meaning the $D_{3d}$ symmetry of the individual RuCl$_3$ layers is unperturbed.

\subsection{II. Thermal Bose Factor and Spectral Weight}

To estimate the temperature dependence of the scattering from purely thermal effects we adopt the following procedure. First, we divide the 295~K spectrum by a factor of $n(\omega, 295~K)+1$. We then multiply this quantity by $n(\omega, T)+1$ to generate the spectra anticipated at other temperatures. Finally, these spectra are integrated as described in the text to obtain the spectral weight (SW). For two-particle scattering, a factor of $(n(\omega, T)+1)^2$ is used instead.

\subsection{III. Fano Resonance Fits}

The Fano effect can occur when a well-defined excitation, in this case an optical phonon, overlaps and also interacts with broad continuum of typically electronic or magnetic origin. These two excitations can interfere, leading to an asymmetric (Fano) resonance. The scattering profile of this resonance as a function of frequency $\omega$ is\cite{thomsen}: 

\begin{equation}\label{fano_lineshape}
f(\omega) = \frac{I_s}{q^2-1}\frac{(q+\epsilon^2)}{1+\epsilon^2}.
\end{equation}

Here $I_s$ is the integrated intensity, $q$ is the Fano asymmetry parameter, and $\epsilon =2(\omega-\omega_o)/\Gamma$ is a reduced frequency determined by the phonon frequency $\omega_o$ and width $\Gamma$. The factor of $q^2-1$ in the denominator is a normalization factor chosen such that the integrated intensity is independent of $q$. 

As detailed in the main text, the low-frequency region is well-described by a Fano parameterization. A fit to a sum of three Lorentzians (one for the continuum and two for the phonons) was also tried, but yielded a consistently worse fit (e.g. the reduced $\chi^2$ was a factor of three larger than that of the Fano parameterization at low temperatures).

In determining the temperature dependence of the 20 meV mode parameters $\omega_o$ and $\Gamma$, we limited our fit to the scattered intensity in the energy range within 2 meV of the phonon peak in order to avoid the complicating effects of the thermal Bose factor. The fact that the phonon frequency $\omega_o$ does not evince any deviations from the anharmonic form is in fact consistent with a continuum density of states that does not vary strongly in the vicinity of $\omega_o$. This further confirms the presence of a broad magnetic continuum rather than well-defined magnon modes that are strongly damped by phonons. This leads to a constant (in energy) $\Gamma$ and a correspondingly small frequency renormalization\cite{thomsen}.

\subsection{IV. Phonon Anharmonicity}
Phonon self-energies are typically determined by anharmonic decay. Decay into a pair of acoustic modes of opposite momenta yields the following expressions for the phonon line width and frequency\cite{PhysRevB.85.012501}:

\begin{equation}\label{phonon_width}
\Gamma(T)=\Gamma_o + A\lbrack 1+2n(\omega_o/2)\rbrack,
\end{equation}
\begin{equation}\label{phonon_frequency}
\omega(T)=\omega_o - B\lbrack 1+2n(\omega_o/2)\rbrack,
\end{equation}

where $\omega_o$ is the bare phonon frequency, $\Gamma_o$ is the zero-temperature line width, $A$ and $B$ are constants, and $n(\frac{\omega_o}{2})$ is the Bose factor for a phonon of energy $\omega_o/2$. Importantly, equations \ref{phonon_width} and \ref{phonon_frequency} are monotonic functions of temperature, varying quasi-linearly for temperatures greater than $\omega_o$ while changing slowly at temperatures less than $\omega_o$. In general, other combinations of acoustic and optical phonons that satisfy energy and momentum conservation can also contribute\cite{PhysRevB.29.2051}. However, these processes lead to a similar monotonic temperature dependence that is almost constant at low temperatures and quasi-linear at high-temperatures. As discussed in the main text, the temperature dependent line width of the 20 meV $E_g$ phonon cannot be described by this expression. In the fits shown in the text, $\Gamma_o$ and $A$ were chosen from a fit to the high temperature linear regime between 160~K and 300~K, while $\omega_o$ was fixed to the 5~K value derived from the Fano fits.


%
\end{document}